\documentclass[aps,prb,reprint,superscriptaddress,longbibliography]{revtex4-2}

\usepackage[T1]{fontenc}
\usepackage{amsmath,amssymb,amsfonts,bm}
\usepackage{graphicx}
\usepackage{booktabs}
\usepackage{multirow}
\usepackage{array}
\usepackage{color}
\usepackage{hyperref}
\usepackage{braket}

\hypersetup{colorlinks=true,linkcolor=blue,citecolor=blue,urlcolor=blue}

\newcommand{\tr}{\operatorname{Tr}}

\newcommand{\Tbar}{\overline{T}}

\newcommand{\class}{\mathcal{C}}
\newcommand{\obs}{\mathcal{O}}

\begin{document}

\title{Disorder signatures in coherent   electronic waveguides}

\author{Alejandro Lopez-Bezanilla}
\affiliation{Theoretical Division, Los Alamos National Laboratory, Los Alamos, New Mexico 87545, USA}

\date{\today}

\begin{abstract}
Microscopic disorder in a coherent conductor is encoded in the magnitude of the transmitted current and in the energy- and channel-resolved structure of the scattering response. 
We develop a conductance-balanced learning framework for identifying microscopic disorder mechanisms from coherent quantum-transport spectra. Using armchair graphene nanoribbons as controlled multichannel tight-binding waveguides, we generate ensembles in which distinct disorder classes occupy the same intervals of integrated transmission, removing the dominant shortcut associated with average conductance. Within this constrained setting, we compare scalar transmission spectra with transmission-eigenvalue spectra, and use supervised classification, scale-normalizing controls, plateau-resolved tests, and principal-component analysis to identify the spectral structures that remain informative. The learned distinctions persist after normalization of the conductance envelope and are strongest in multichannel energy windows, where mode mixing and channel redistribution shape the scattering response. Principal components provide interpretable transport coordinates whose loadings identify the energy and eigenchannel sectors responsible for the dominant spectral deformations. The results establish a controlled AI-assisted protocol for learning disorder signatures from nonlinear quantum-scattering data.
\end{abstract}

\maketitle

\section{Introduction}

Transmission through a structured medium is a boundary probe of the internal scattering landscape. This perspective is common to optical speckle, wavefront shaping, acoustic and microwave scattering, elastic-wave propagation, and coherent electronic conduction: microscopic inhomogeneity inside the medium is converted into an input--output response measured at external channels or detectors. The forward problem is well posed once the Hamiltonian or wave operator, contacts, and boundary conditions are known. The inverse problem is more constrained by physics. One asks which properties of a high-dimensional internal structure can be inferred from a finite set of transmitted observables. In coherent multiple-scattering systems the map is nonlinear and generally many-to-one because interference, resonances, evanescent propagation, and mode conversion combine contributions from many paths before any external observable is measured~\cite{ColtonKress2013,RotterGigan2017,Mosk2012,Donoho2006,Duarte2008,Shapiro2008,Bromberg2009}.

Mesoscopic electronic transport provides a sharp realization of this inverse problem. The scattering medium is a Hamiltonian, and the measured response is encoded in the scattering matrix connecting incoming and outgoing lead modes. Within the Landauer--B\"uttiker picture, conductance is determined by transmission probabilities; in the Green-function formulation those probabilities are nonlinear functionals of the central-region Hamiltonian and of the lead self-energies~\cite{Landauer1957,Landauer1970,Buttiker1986,Caroli1971,FisherLee1981,Datta1995,NazarovBlanter2009}. A transmission spectrum should therefore be interpreted as a coherent scattering fingerprint rather than as a direct image of the disorder landscape. The same microscopic perturbation can affect a spectrum through phase shifts, quasi-bound states, intersubband scattering, intervalley mixing, and band-edge modifications. Conversely, distinct microscopic disorder configurations can generate very similar two-terminal spectra.

The transmission eigenvalues provide a natural language for this problem. In a multichannel conductor the scalar transmission gives the total transmitted weight, whereas the eigenvalues of the transmission matrix determine how that weight is distributed among eigenchannels. This distinction is central in mesoscopic transport: transmission eigenvalue distributions underlie conductance fluctuations, shot noise, full-counting-statistics observables, and localization theory in quasi-one-dimensional conductors~\cite{Dubois2010,MelloPereyraKumar1988,BeenakkerRMP1997,NazarovBlanter2009}. For the present purpose, the eigenvalue spectrum also separates physically distinct types of conductance loss. Smooth scalar potentials, random hopping, dopants, vacancies, and edge disorder can produce comparable reductions in total transmission while suppressing or redistributing channels in different ways.

\begin{figure*}[htp]
    \centering
    \includegraphics[width=0.95\linewidth]{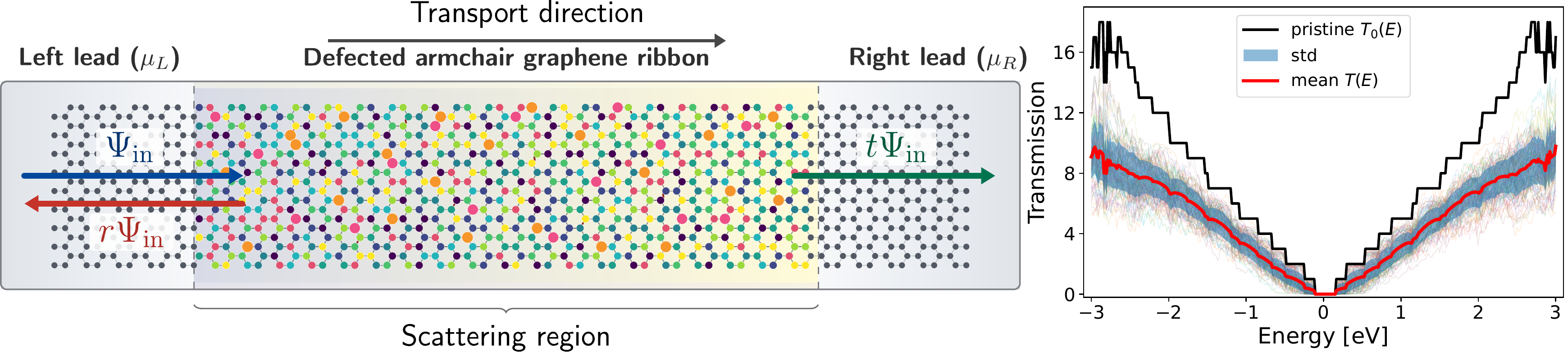}
\caption{
Schematic transport setup and transmission statistics for a defected armchair graphene ribbon.
The left panel shows a disordered scattering region connected to left and right defect-free leads with electrochemical potentials $\mu_L$ and $\mu_R$. An incoming mode $\Psi_{\rm in}$ from the left lead is partially reflected, $r\Psi_{\rm in}$, and partially transmitted, $t\Psi_{\rm in}$, through the defected channel; colored sites indicate spatial disorder in the ribbon. The right panel shows the corresponding ensemble transmission $T(E)$ as a function of energy. Thin curves denote individual disorder realizations, the red curve is the ensemble-averaged transmission $\langle T(E)\rangle$, the blue band indicates the standard deviation, and the black staircase curve is the pristine reference transmission $T_0(E)$.
}
    \label{fig0}
\end{figure*}

Graphene nanoribbons are useful as a controlled platform for this question. A nearest-neighbor $\pi$-orbital model gives a sparse Hamiltonian with well-defined leads, sublattice structure, transverse quantization, edge sensitivity, and several physically interpretable disorder channels~\cite{Nakada1996,Son2006,BreyFertig2006,CastroNeto2009,Kwant2014}. Scalar onsite disorder models electrostatic inhomogeneity or charge puddles. Hopping disorder modifies the covalent network and is naturally associated with bond-length fluctuations, strain-like perturbations, or local chemistry. Dopants introduce localized onsite scattering centers and can generate resonant spectral features. Vacancies and edge roughness disrupt connectivity and are particularly effective in narrow nanoribbons, where boundary conditions control the transverse modes and can promote localization or strong sample-to-sample variability~\cite{Gallagher2010,Areshkin2007,Mucciolo2009,Cresti2011}.

The objective of this work is to determine which aspects of these microscopic disorder mechanisms remain visible in coherent transport after the most obvious scalar cue, the average transmission, has been removed. The analysis targets the disorder-dependent structure that survives in the scattering response, rather than a one-to-one reconstruction of the microscopic disorder realization. The task is to infer the disorder mechanism and associated statistical descriptors from energy-resolved transport observables, while accepting that a unique reconstruction of the disorder field is generally unavailable from two-terminal data alone. This distinction is essential: a high classification accuracy in an unconstrained data set could merely indicate that one disorder class is stronger or more opaque than another. A physically meaningful test must compare disorder classes at fixed integrated transmission.

Here we construct conductance-balanced ensembles of disordered armchair graphene nanoribbons. Each disorder class is sampled so that it populates the same windows of energy-averaged transmission. We then inquire whether the remaining spectral structure distinguishes the underlying Hamiltonian perturbation. The analysis compares the scalar transmission $T(E)$ with the transmission-eigenvalue spectrum $\{\tau_n(E)\}$, applies hard normalizations that remove conductance scale, and interprets the residual disorder information using plateau-resolved spectral tests, channel-participation descriptors, and principal-component transport coordinates. This organization keeps the learning machinery deliberately simple, so that the central claim concerns the physics of the transport observables: class-dependent disorder information persists in spectral shape and eigenchannel redistribution even when average conductance is controlled.

\section{Tight-binding model of the ribbon waveguide}
\label{sec:model}

We study an armchair graphene ribbon connected to two semi-infinite pristine leads. The scattering region has width $N=36$ armchair dimer lines and length $L=56$ armchair unit cells. The purpose of the model is to provide a reproducible multichannel tight-binding waveguide with graphene lattice connectivity. The conclusions are therefore phrased in terms of coherent waveguide transport rather than as a claim about a single experimentally realized graphene device.

The system is partitioned into left lead, central scattering region, and right lead, as shown schematically in Fig.~\ref{fig0}. The semi-infinite leads are ballistic and have the same defect-free lattice structure as the clean ribbon. The contacts are thus mode matched, such that elastic backscattering is generated by the disorder inside the finite central region rather than by lead mismatch. This choice isolates the intrinsic scattering signatures of the disorder classes and avoids conflating them with contact-induced reflections.

The pristine Hamiltonian is
\begin{equation}
H_0=
\sum_i \epsilon_i^{(0)} c_i^\dagger c_i
-
\sum_{\langle i,j\rangle}t_{cc}
\left(c_i^\dagger c_j+c_j^\dagger c_i\right),
\label{eq:H0}
\end{equation}
where $c_i^\dagger$ creates an electron in a $p_z$ orbital on site $i$, and $\langle i,j\rangle$ denotes nearest-neighbor pairs. The onsite energy is $\epsilon_i^{(0)}=0$, and the nearest-neighbor carbon-carbon hopping is $t_{cc}=2.7\,\mathrm{eV}$. Boundary atoms may carry the same fixed edge onsite shift in both the scattering region and the leads to mimic effective passivation; because this term is included in the reference Hamiltonian, it acts as a fixed model parameter rather than sample-dependent disorder. The same pristine lead Hamiltonian is used for all samples.

A disordered realization is written as $ H_s=H_0+\delta H_{\rm dis}$,
with
\begin{equation}
\delta H_{\rm dis}=
\sum_i V_i c_i^\dagger c_i
-
\sum_{\langle i,j\rangle}\delta t_{ij}
\left(c_i^\dagger c_j+c_j^\dagger c_i\right)
+\delta H_{\rm def}.
\label{eq:disorder_general}
\end{equation}
Here $V_i$ is a scalar onsite perturbation, $\delta t_{ij}$ is a bond perturbation, and $\delta H_{\rm def}$ represents discrete defects such as dopants, vacancies, or edge modifications. This decomposition is physically important. Onsite disorder shifts local orbital energies and primarily acts as an electrostatic perturbation. Hopping disorder changes the connectivity and velocity structure of the lattice Hamiltonian. Vacancies and edge defects remove or strongly modify sites and therefore create short-range scatterers that can mix transverse modes and valleys more efficiently than a smooth scalar field.

\subsection{Disorder classes}
\label{subsec:disorder_classes}

\begin{figure}
    \centering
    \includegraphics[width=0.95\linewidth]{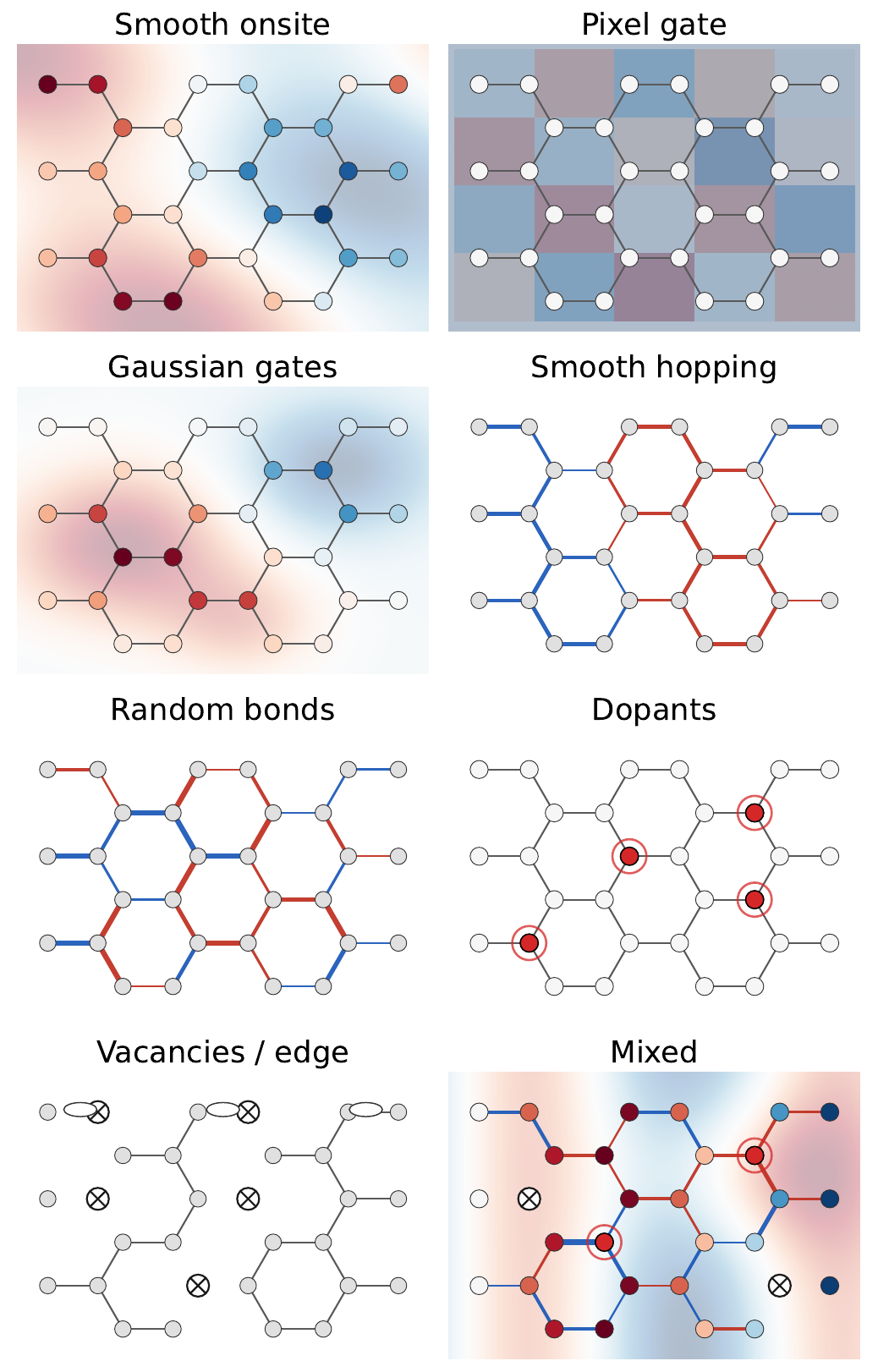}
    \caption{Catalog of disorder mechanisms used in the conductance-balanced transport ensemble. The panels schematically represent scalar onsite perturbations, gate-defined potentials, hopping-amplitude disorder, localized dopants, vacancies and edge roughness, and mixed perturbations. The figure provides a schematic overview of the Hamiltonian modifications applied to the finite scattering region, rather than a literal microscopic realization from the dataset; the semi-infinite leads remain pristine.}
    \label{fig1} 
\end{figure}

In our benchmark, eight disorder families are used. Figure~\ref{fig1} displays the different types of defects considered in this work. They are chosen to span several qualitatively different Hamiltonian perturbations while retaining a common geometry, lead structure, and energy grid. The classes are defined at the Hamiltonian level; their strength parameters are sampled over prescribed ranges before the conductance-balancing selection described in Sec.~\ref{sec:fair}.

\emph{Smooth onsite disorder.}  The onsite potential $V_i=V(\bm r_i)$ is a correlated smooth field generated from a finite set of long-wavelength Fourier components. This class models electrostatic inhomogeneity, smooth substrate disorder, or remote-charge puddles. Because the perturbation varies slowly on the lattice scale, it mainly produces phase accumulation, gradual mode-dependent reflection, and broad plateau distortions. It is comparatively inefficient at producing atomically sharp intervalley scattering.

\emph{Pixelated gate disorder.}  A coarse real-valued grid is sampled and interpolated onto the atomic sites. This produces a scalar gate landscape with a controlled spatial resolution. It is rougher than the Fourier field while remaining onsite in character. This class bridges smooth electrostatic disorder and high-dimensional gate patterns of the kind that could be imposed by structured local gates or electrostatic environments.

\emph{Gaussian gate disorder.}  The potential is a sum of localized Gaussian barriers or wells,
$
V(\bm r)=\sum_{k=1}^{K}A_k
\exp\left[-\frac{|\bm r-\bm R_k|^2}{2\sigma_k^2}\right].
\label{eq:gaussian_gate}
$
The amplitudes, centers, widths, and number of Gaussians vary across realizations. This class represents localized puddles, gate-defined barriers, or smooth impurity clusters. Compared with the Fourier and pixel classes, it can introduce more localized resonant structure while preserving the scalar onsite nature of the perturbation.

\emph{Smooth hopping disorder.}  The nearest-neighbor hoppings are modified by smooth bond-direction-dependent fields,
$
t_{ij}=t_{cc}\left[1+\eta_{d(i,j)}(\bm r_{ij})\right],
\label{eq:hopping_smooth}
$
where $d(i,j)$ labels the bond orientation and $\bm r_{ij}$ is the bond center. Smooth hopping disorder emulates strain-like or chemistry-induced bond modulation. It changes the kinetic part of the Hamiltonian, and therefore it can perturb velocities, subband dispersions, and mode mixing in a way that differs from scalar onsite disorder.

\emph{Random bond disorder.}  The hopping amplitudes are perturbed on shorter length scales, producing a bond network with stronger local randomness. This class is a controlled representation of lattice-scale hopping fluctuations. Because the perturbation enters through bonds, it can broaden subbands, alter current pathways, and mix transverse channels even when the total transmission is constrained to match scalar-potential samples.

\emph{Dopants.}  Substitutional dopants are represented by a random set of sites carrying onsite shifts. A dopant realization is characterized by a dopant density and by the distribution of onsite strengths. The scattering is localized in real space and can therefore produce Fano-like resonances, antiresonances, and particle-hole asymmetry that differ from the broad deformations associated with smooth scalar gates.

\emph{Vacancies and edge disorder.}  Vacancies are introduced by removing sites or equivalently decoupling them from the tight-binding network. Edge roughness is generated by removing or modifying atoms near the ribbon boundaries. This class is important in narrow waveguides because transverse quantization and boundary conditions determine the propagating modes. Edge perturbations can strongly affect modes with large boundary weight, while bulk vacancies can create localized states and transmission zeros.

\emph{Mixed disorder.}  The mixed class contains multiple disorder ingredients in the same sample. It is included as a deliberately nonideal class: many real devices contain overlapping perturbations rather than a single clean disorder mechanism. Its role is to test whether the transport observables identify spectra produced by superposed mechanisms or whether they assign such spectra to the nearest single-mechanism class in transport space.

\subsection{Physical expectations from the disorder catalog}
\label{subsec:physical_expectations}

The disorder classes can be organized by three physical attributes: the Hamiltonian sector they perturb, their real-space correlation length, and the symmetries they preserve or break at the lattice scale. This catalog is useful because the classifier is expected to learn physically meaningful distinctions rather than arbitrary labels. It should separate classes only when the associated perturbations produce distinct scattering processes in the available modes. A smooth onsite field and a smooth Gaussian gate both enter as scalar potentials. Their strongest signatures should therefore appear as energy-dependent phase shifts, changes in local carrier density, and broad distortions of conductance plateaus. Their ability to mix valleys or strongly couple transverse modes is limited when their spatial Fourier content remains small compared with reciprocal-lattice momenta. This expectation explains why scalar-potential-like classes are the most likely to overlap after conductance balancing.

Short-range onsite disorder and dopants have a different scattering character. They still perturb the onsite sector, yet their spatial localization introduces large momentum components and can couple states that a smooth field would leave nearly independent. Localized onsite shifts can also create quasi-bound states whose interference with propagating modes generates narrow suppressions of $T(E)$. In transport language, these appear as resonance and antiresonance features superimposed on the broader subband structure. Such features can survive conductance balancing because their energy positions and widths are governed by factors beyond the integrated transmission. They depend on the microscopic arrangement of dopants, their sublattice positions, and their coupling to the propagating eigenchannels.

Hopping disorder perturbs the kinetic part of the Hamiltonian. In the long-wavelength Dirac description of graphene, slowly varying bond modulations can be interpreted in terms of effective gauge-field-like perturbations and local velocity changes, while atomically sharp bond randomness has substantial intervalley and intersubband content. In the tight-binding waveguide used here, this distinction appears through the transmission eigenvalues: a hopping perturbation can redistribute current across channel ranks even when the scalar transmission is forced to match an onsite-disorder sample. The same value of $T(E)$ may arise from nearly uniform attenuation of many eigenchannels or from stronger mixing that converts a few open channels into several partially transmitting channels.

Vacancies and edge roughness form the most singular class in the catalog. A vacancy changes the graph on which the tight-binding Hamiltonian is defined; edge roughness changes the boundary conditions that quantize transverse modes. These perturbations are expected to be especially visible in narrow nanoribbons because a sizable fraction of the wave-function weight can reside near the edges, and because the separation between transverse subbands is large enough for mode-dependent scattering to be resolved over the energy grid. In a noninteracting nearest-neighbor model, the calculation excludes magnetic reconstruction around vacancies or zigzag segments. It isolates the single-particle scattering consequence of connectivity loss and boundary modification.

This physical catalog also clarifies the role of the mixed class. Mixed disorder provides a critical test of whether transport spectra can be described as linear combinations of single-mechanism signatures or whether interference among perturbations produces distinct regions of transport space. Because coherent scattering is nonlinear in the Hamiltonian perturbation, the spectrum of a mixed sample is generally more than a weighted sum of spectra from its ingredients. The mixed class is therefore expected to overlap with several single-mechanism classes while still carrying information about the combined scattering landscape.

\section{Transport observables and metrological formulation}
\label{sec:transport}

For each Hamiltonian $H_s$ we compute coherent two-terminal transport on a fixed energy grid $E\in[-3,3]~\mathrm{eV}$ with $N_E=401$ points using Kwant's scattering-matrix solver. Kwant returns the dimensionless transmission associated with the degrees of freedom explicitly included in the tight-binding Hamiltonian. In the spinless model used here, the corresponding spin-degenerate linear-response conductance is
\begin{equation}
G(E_F)=\frac{2e^2}{h}T(E_F),
\label{eq:landauer_conductance}
\end{equation}
where $T(E_F)$ is the total transmission at the Fermi energy. The full energy dependence $T(E)$ is used in the present analysis rather than a single Fermi-level value.

For each energy $E$, Kwant constructs the propagating modes of the semi-infinite leads and solves the stationary scattering problem for the finite disordered region attached to those leads. The scattering matrix relates incoming and outgoing propagating modes and is expressed in a current-normalized lead-mode basis. In block form,
\begin{equation}
S(E)=
\begin{pmatrix}
r(E) & t'(E) \\
t(E) & r'(E)
\end{pmatrix},
\label{eq:scattering_matrix}
\end{equation}
where $r$ and $r'$ are reflection matrices and $t$ and $t'$ are transmission matrices. The block $t(E)$ maps incoming modes in the left lead into outgoing modes in the right lead. The scalar two-terminal transmission returned by the scattering matrix is
\begin{equation}
T(E)=\tr\left[t^\dagger(E)t(E)\right].
\label{eq:T_from_t}
\end{equation}
Equivalently, the same coherent elastic transmission can be written in the Green-function or Caroli form,
\begin{equation}
T(E)=
\tr\left[
\Gamma_L(E)G^r(E)\Gamma_R(E)G^a(E)
\right],
\label{eq:caroli_formula}
\end{equation}
with
\begin{equation}
G^r(E)=
\left[(E+i0^+)I-H_s-\Sigma_L^r(E)-\Sigma_R^r(E)\right]^{-1}
\label{eq:retarded_green}
\end{equation}
and $G^a=(G^r)^\dagger$. The lead broadenings are $\Gamma_\alpha(E)=i[\Sigma_\alpha^r(E)-\Sigma_\alpha^a(E)]$ for $\alpha=L,R$. Equations~\eqref{eq:T_from_t} and \eqref{eq:caroli_formula} are equivalent forms of the Landauer--Caroli transmission for coherent elastic transport; the scattering-matrix expression in Eq.~\eqref{eq:T_from_t} is the numerical route used in the present calculations.

The transmission eigenvalues are the eigenvalues of the positive semidefinite matrix
$\mathcal{T}(E)=t^\dagger(E)t(E)$,
so that
\begin{equation}
\mathcal{T}(E)|\phi_n^{\rm in}(E)\rangle
=
\tau_n(E)|\phi_n^{\rm in}(E)\rangle,
\qquad
0\leq \tau_n(E)\leq 1.
\label{eq:transmission_eigenproblem}
\end{equation}
The vectors $|\phi_n^{\rm in}\rangle$ define incoming transmission eigenchannels, which are coherent superpositions of lead modes that diagonalize the transmission problem at fixed energy. A singular-value decomposition,
\begin{equation}
t(E)=U(E)\,
\mathrm{diag}\left(\sqrt{\tau_1(E)},\sqrt{\tau_2(E)},\ldots\right)
V^\dagger(E),
\label{eq:svd_transmission}
\end{equation}
makes the channel interpretation explicit. The total transmission is recovered as
\begin{equation}
T(E)=\sum_n\tau_n(E).
\label{eq:T_sum_tau}
\end{equation}
Thus $T(E)$ reports the total transmitted weight, while $\{\tau_n(E)\}$ records how that weight is partitioned among eigenchannels. Two samples with equal $T(E)$ can differ sharply: one may transmit through a small number of nearly open channels, whereas another may distribute current over many partially transmitting channels.

For numerical learning, the eigenvalues are sorted by magnitude at each energy and stored as an energy-dependent spectrum. Closed or absent channels are represented by zero-valued entries when a fixed feature dimension is required. This sorting and padding are postprocessing conventions applied to the Kwant scattering-matrix output. The index $n$ therefore labels an eigenvalue rank rather than a fixed transverse lead mode over the entire energy range. This convention retains the information needed to analyze disorder-induced redistribution among transmission eigenchannels while providing a uniform feature representation.

The direct microscopic inverse problem would seek a map
$T(E)\longrightarrow \delta H_{\rm dis},$
Preliminary tests with atom-resolved onsite fields and coarse pixel fields show the expected nonuniqueness: samples close in spectral distance can remain widely separated in microscopic disorder space, and deterministic inverse networks recover broad trends more reliably than the original field. This behavior reflects the many-to-one character of the coherent scattering map.

The more appropriate formulation is metrological. We ask whether a transport observable determines a disorder class and a set of statistical descriptors,
\begin{equation}
\obs(E)\longrightarrow \left(\class,\bm\theta_{\rm dis}\right),
\label{eq:metrology_map}
\end{equation}
where $\obs$ denotes $T(E)$, $\{\tau_n(E)\}$, or their concatenation. The descriptor vector may include disorder strengths, hopping-amplitude scales, dopant counts, vacancy fractions, and the induced standard deviations of onsite and hopping fields. This formulation recognizes that two-terminal data generally lack sufficient information to identify the microscopic disorder realization. It asks instead which coarse Hamiltonian information survives the projection from disorder space to transport space.

The eigenvalue hierarchy also connects the present analysis to experimentally motivated observables. The conductance depends on the first moment of the eigenvalue distribution, $\sum_n\tau_n$. The zero-temperature shot-noise power contains the second combination $\sum_n\tau_n(1-\tau_n)$, often expressed through the Fano factor
\begin{equation}
F(E)=\frac{\sum_n\tau_n(E)[1-\tau_n(E)]}{\sum_n\tau_n(E)}.
\label{eq:fano_factor}
\end{equation}
Higher current cumulants probe higher moments. The full transmission-eigenvalue spectrum used in this numerical study is therefore an idealized upper-information observable within coherent two-terminal transport. Comparing $T(E)$ with $\{\tau_n(E)\}$ establishes how much additional disorder information is available when the distribution over channels is resolved beyond its first moment.

Several compact descriptors are useful for interpreting the channel-resolved spectra. The effective number of transmitting channels is
\begin{equation}
N_{\rm eff}(E)=
\frac{\left[\sum_n\tau_n(E)\right]^2}{\sum_n\tau_n^2(E)},
\label{eq:Neff_definition}
\end{equation}
which is a participation ratio in eigenchannel space. It approaches unity when one eigenchannel dominates and increases when several eigenchannels carry comparable current. A normalized eigenchannel entropy is defined as
\begin{equation}
\begin{aligned}
S_\tau(E)&=-\frac{1}{\ln N_{\rm prop}(E)}
\sum_{n=1}^{N_{\rm prop}(E)}p_n(E)\ln p_n(E),\\
p_n(E)&=\frac{\tau_n(E)}{\sum_m\tau_m(E)}.
\end{aligned}
\label{eq:entropy_tau_definition}
\end{equation}
Here $N_{\rm prop}(E)$ is the number of propagating channels retained at energy $E$, and zero eigenvalues are excluded from the sum. When $T(E)=0$ or $N_{\rm prop}(E)\leq1$, the entropy is set to zero by convention. These descriptors complement the full spectra by providing interpretable summaries of mode participation and channel redistribution.

\section{Conductance-balanced benchmark construction}
\label{sec:fair}

A meaningful disorder-metrology benchmark must separate microscopic scattering fingerprints from the trivial fact that different disorder mechanisms may have different average scattering strengths. Without such control, a classifier could identify a disorder class by using only the energy-averaged transmission. That result would mainly establish an opacity hierarchy among the sampled disorder classes. It would give little evidence that the spectra contain mechanism-specific structure beyond total conductance loss.

For each sample we compute
\begin{equation}
\Tbar=\frac{1}{N_E}\sum_{\ell=1}^{N_E}T(E_\ell),
\label{eq:Tbar_balancing}
\end{equation}
where the sum runs over the 401 energies in the interval $[-3,3]$ eV. We then perform acceptance sampling in windows of $\Tbar$.
The production data set was generated with a target of 400 samples per disorder class and conductance bin over the interval \(4.0<\overline{T}<5.75\), divided into three equal-width bins. Since a few class--bin cells fell short of the target, the final conductance-balanced subset was constructed by retaining 307 samples from every class--bin cell, giving \(8\times3\times307=7368\) spectra.

Conductance balancing removes a dominant scalar variable while preserving the physical constraints imposed by the band structure. Samples in the same $\Tbar$ bin can have different distributions of transmission over energy because the average in Eq.~\eqref{eq:Tbar_balancing} is insensitive to where transmission is lost. A disorder realization that weakly suppresses many plateaus and another that produces localized antiresonances can have the same $\Tbar$. Similarly, two samples can have the same scalar spectrum while displaying different channel participation. The benchmark is therefore intentionally stringent without being artificially impossible: it suppresses the most trivial cue while preserving the spectral degrees of freedom that coherent scattering theory predicts should remain mechanism dependent.

The use of three bins rather than a single narrow window also matters. A single extremely narrow conductance interval would reduce the data set and could overemphasize accidental spectral degeneracies. Three balanced windows retain a range of transparencies while enforcing the same transmission distribution for every class. The classification task therefore asks whether each class occupies a reproducible region of transport space across several conductance levels, rather than whether a single carefully selected opacity slice can be separated.

This construction fixes the role of $\Tbar$. The model can still use the detailed energy dependence of $T(E)$ and $\{\tau_n(E)\}$, including plateau distortions, particle-hole asymmetry, resonances, antiresonances, and mode-dependent suppression. The task requires more than a simple rule that assigns one disorder class to low average transmission and another to high average transmission. In the balanced subset, class-averaged values of $\Tbar$ lie in a narrow range, from approximately 4.80 for the pixel-gate class to approximately 4.94 for the mixed and vacancy/edge classes. The remaining differences are small compared with the class separation in unconstrained ensembles and are further tested using area-normalized and standardized spectra in Sec.~\ref{subsec:shape_controls}.

\section{Learning protocol and control transformations}
\label{sec:ai}

The learning task is supervised disorder metrology. Each sample is represented by one of three transport observables,
\begin{align}
\bm x_T &= \left[T(E_1),\ldots,T(E_{N_E})\right], \label{eq:features_T}\\
\bm x_\tau &= \left[\tau_1(E_1),\ldots,\tau_{N_{\rm max}}(E_1),\ldots,
\tau_{N_{\rm max}}(E_{N_E})\right], \label{eq:features_tau}\\
\bm x_{T+\tau} &= \bm x_T\oplus\bm x_\tau . \label{eq:features_combined}
\end{align}
The scalar input has 401 components. The eigenchannel representation stores $N_{\rm max}=32$ eigenvalues at each energy in the present geometry, giving 12832 components. The combined representation has 13233 components.

Random-forest classifiers and regressors are trained on principal-component analysis (PCA)-compressed features, while neural-network baselines use standardized inputs. These models are intentionally conventional. Their role is to determine whether the chosen transport observable retains recoverable information about the disorder mechanism and its descriptors. The central interpretation comes from comparisons among observables and from controlled transformations of the spectra, rather than from the architecture itself.

The categorical target is the disorder class $\class$. The continuous targets include strengths of onsite, pixel-gate, Gaussian-gate, hopping, and random-bond disorder. And also dopant density, vacancy fraction, and induced standard deviations of onsite and hopping fields. We report class accuracy and the mean coefficient of determination $R^2$ over these continuous descriptors. These numerical scores are used as diagnostics of recoverability in transport space.

Two scale-removing transformations are central. In the area-normalized representation, each spectrum is divided by its integrated spectral weight. In the per-spectrum standardized representation, each spectrum is centered and divided by its own feature-axis standard deviation. For the scalar $T(E)$ input, we also use a finite-difference representation along the energy axis. These controls suppress simple spectral cues, such as the overall conductance scale, and test whether class information remains in the shape of the response. Robust performance after area normalization or standardization indicates that the model is using correlated deformations of the spectrum rather than a residual offset in conductance scale.

\begin{figure*}[htp]
    \centering
    \includegraphics[width=0.995\linewidth]{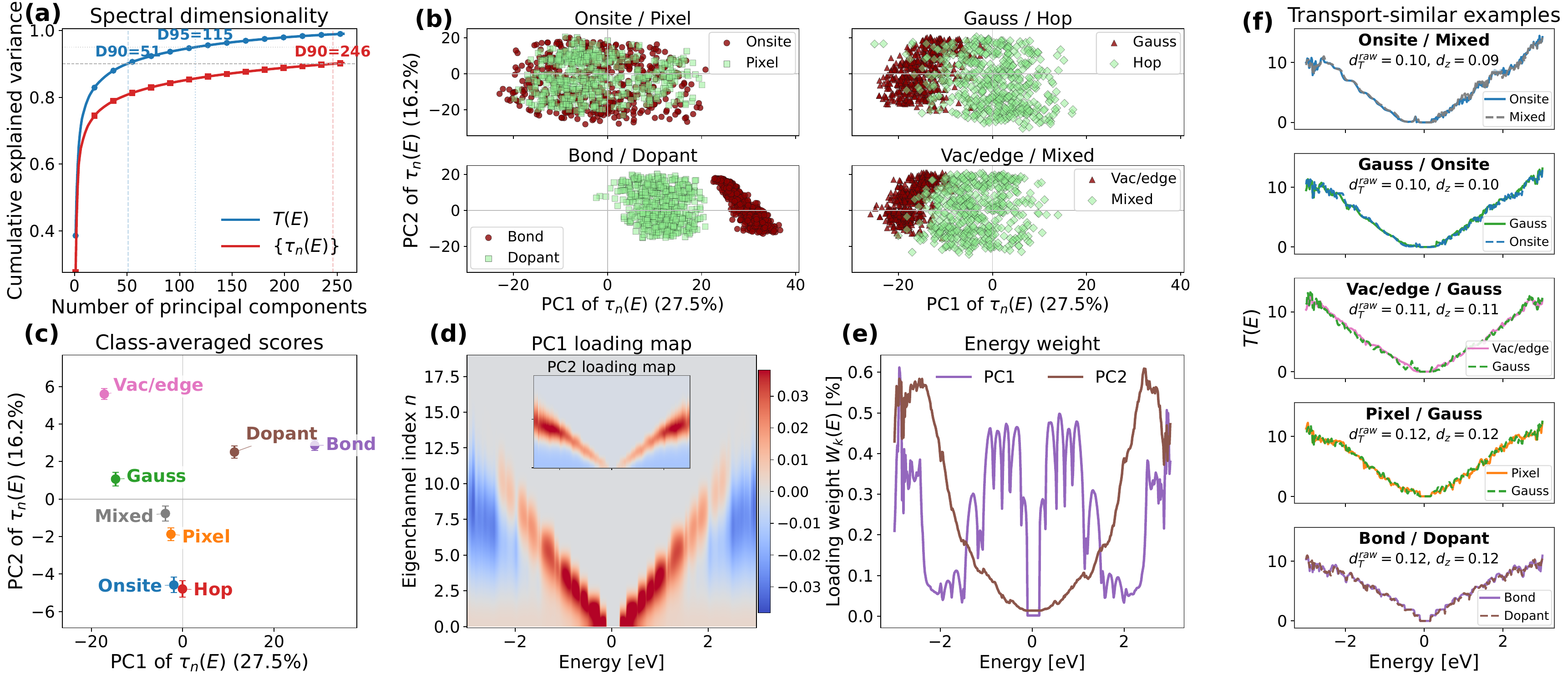}
    \caption{
Principal-component analysis of the conductance-balanced transport ensemble.
(a) Cumulative explained variance for the scalar transmission $T(E)$ and the transmission-eigenvalue spectrum $\{\tau_n(E)\}$.
(b) Individual disorder realizations projected onto the first two principal components of $\{\tau_n(E)\}$, showing the overlap and separation of representative classes in transport space.
(c) Class-averaged PC1-PC2 scores for $\{\tau_n(E)\}$; error bars denote the standard error of the mean.
(d,e) Loading maps of the first two principal components, $u_1(E,n)$ and $u_2(E,n)$, resolved by energy and sorted transmission-eigenvalue index $n$.
(f) Energy-resolved loading weights $W_\alpha(E)=\sum_n u_\alpha^2(E,n)$, indicating which spectral regions dominate the leading PCA coordinates.
The principal components provide empirical transport coordinates for the dominant spectral deformations of the conductance-balanced ensemble.
}
    \label{fig:pca_transport_coordinates}
\end{figure*}

\section{Results}
\label{sec:results}

\subsection{Spectral dimensionality and PCA transport coordinates}
\label{subsec:pca_transport_coordinates}

\begin{table}[t]
\caption{PCA dimensionality of the conductance-balanced production subset. $D_p$ is the number of principal components required to explain $p\%$ of the variance. The table is used as an information-content diagnostic rather than as the primary performance result.}
\label{tab:pca_revised}
\begin{ruledtabular}
\begin{tabular}{lccc}
Observable & $D_{90}$ & $D_{95}$ & $D_{99}$ \\
\hline
$T(E)$ & 22 & 66 & 215 \\
$\{\tau_n(E)\}$ & 541 & 1053 & 2180 \\
$T(E)+\{\tau_n(E)\}$ & 492 & 995 & 2130 \\
\end{tabular}
\end{ruledtabular}
\end{table}

A useful first diagnostic is the number of independent spectral directions required to describe the conductance-balanced ensemble. We use PCA as an interpretable linear decomposition and as a controlled dimensionality-reduction step for the random-forest baseline. Rather than using PCA as evidence for a microscopic mechanism by itself, its purpose is to identify whether the spectra vary along a few trivial transmission directions or occupy a higher-dimensional manifold of transport deformations.

For a set of standardized transport features $X_i$, PCA writes
\begin{equation}
X_i\simeq \bar X + \sum_\alpha z_{i\alpha}u_\alpha,
\label{eq:pca_decomposition}
\end{equation}
where $u_\alpha$ is the $\alpha$-th principal component and $z_{i\alpha}$ is the score of sample $i$ along that component. For the scalar transmission, $X_i$ is the function $T_i(E)$. For the eigenchannel representation, $X_i$ is the two-dimensional object $\tau_{i,n}(E)$, flattened before PCA and reshaped into energy and eigenvalue-rank indices for interpretation.

Table~\ref{tab:pca_revised} gives the number $D_p$ of principal components required to explain $p\%$ of the variance in the balanced subset. For example, for the scalar transmission, 66 components are required to capture 95\% of the variance. The eigenchannel representation requires 1053 components to reach the same threshold. This dimensional increase has a direct physical interpretation: after average conductance has been controlled, the remaining variability is distributed across mode-opening distortions, resonant structures, particle-hole asymmetries, and channel-redistribution patterns. Thus, the ensemble requires more than a single opacity coordinate.

Figure~\ref{fig:pca_transport_coordinates} summarizes the PCA transport coordinates. Panel (a) shows cumulative explained variance for $T(E)$ and $\{\tau_n(E)\}$. Panels (b) and (c) show how disorder mechanisms are organized in the leading PCA plane of the eigenchannel spectra. The class clouds overlap, (b), is consistent with transport-similar microscopic perturbations, yet the class centroids, (c), occupy reproducibly different regions of the same plane. The scores therefore act as empirical transport coordinates: without reconstructing the disorder field, they reveal how disorder mechanisms are embedded in the dominant directions of spectral variability.

\begin{table}[t]
\caption{Main supervised-learning scores for the conductance-balanced data set. Accuracy refers to disorder-class classification. The regression score is the mean \(R^2\) over continuous disorder descriptors.}
\label{tab:main_scores}
\begin{ruledtabular}
\begin{tabular}{lcc}
Input representation & Accuracy & Mean \(R^2\) \\
\hline
\(T(E)\) & 0.887 & 0.614 \\
\(\{\tau_n(E)\}\) & 0.922 & 0.731 \\
\(T(E)+\{\tau_n(E)\}\) & 0.923 & 0.735 \\
\end{tabular}
\end{ruledtabular}
\end{table}

The loading maps in panels (d) and (e) connect these coordinates to the multichannel scattering structure. The sign of a principal component is arbitrary; the spatial pattern of positive and negative loading over $(E,n)$ is meaningful because it identifies which parts of the transmission-eigenvalue spectrum are enhanced or suppressed as the PCA score changes. The leading deformations are concentrated in selected energy and eigenchannel sectors rather than being uniformly distributed across the band. The energy-resolved loading weight $W_\alpha(E)=\sum_n u_\alpha^2(E,n)$
shows that the multichannel regions carry substantial weight in the dominant PCA coordinates. This observation anticipates the energy-window analysis below, where higher-energy sectors are more diagnostic than the central low-energy sector.

\subsection{Transport distinguishability}
\label{subsec:distinguishability}

We summarize the class-identification results through transport distinguishability maps constructed from row-normalized confusion matrices. Let \(C_{ij}\) denote the number of test spectra whose actual disorder mechanism is \(i\) and whose predicted mechanism is \(j\). We plot
\begin{equation}
M_{ij}
=
\frac{C_{ij}}{\sum_j C_{ij}},
\label{eq:row_normalized_confusion}
\end{equation}
so that each row gives the conditional probability \(P(\widehat{c}=j|c=i)\) of assigning a spectrum from class \(i\) to class \(j\). In this representation, large diagonal entries indicate mechanisms whose spectra are assigned to their own class, whereas off-diagonal entries identify pairs of mechanisms that generate similar transport responses. Because the data set is conductance balanced, the off-diagonal structure probes similarity in the energy- and channel-resolved scattering response at comparable integrated transmission, rather than differences in the overall conductance scale.

For the balanced subset, the random-forest baseline trained on \(T(E)\) reaches an accuracy of 0.887 and a mean regression \(R^2\) of 0.614. The eigenchannel representation gives 0.922 and 0.731, respectively, while the combined representation gives 0.923 and 0.735. These metrics are summarized in Table~\ref{tab:main_scores}. The small gain obtained by adding \(T(E)\) to \(\{\tau_n(E)\}\) is consistent with the identity \(T(E)=\sum_n \tau_n(E)\): the scalar transmission is contained in the eigenvalue spectrum, while the latter also resolves how the transmitted current is distributed over channel ranks.

\begin{figure}[t]
\includegraphics[width=\linewidth]{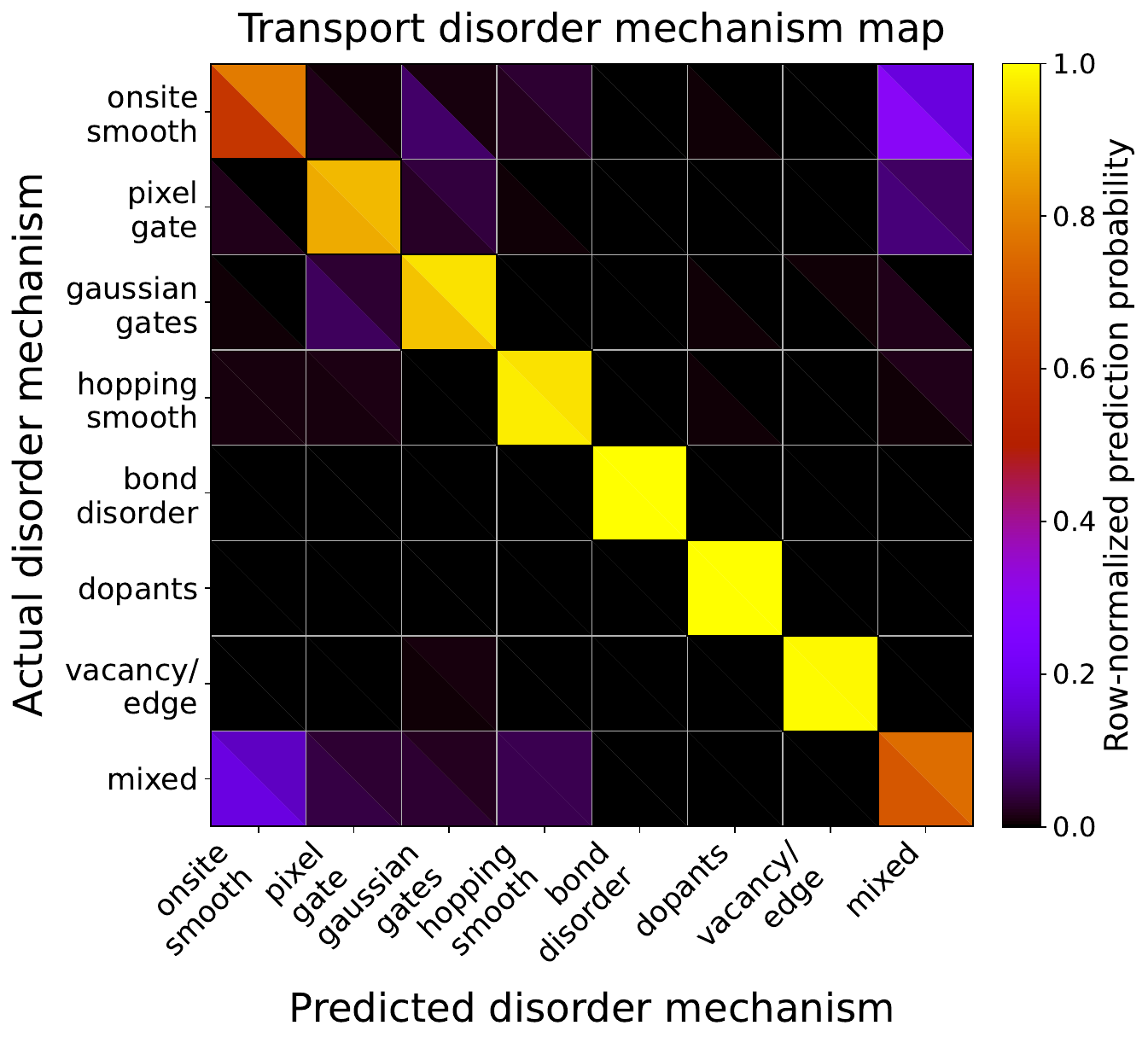}
\caption{
Transport distinguishability map for the conductance-balanced disorder ensemble.
Rows denote the actual disorder mechanism and columns the predicted mechanism. The plotted entries are row-normalized prediction probabilities, Eq.~\eqref{eq:row_normalized_confusion}. Each square is split by input representation: lower-left triangles use the total transmission \(T(E)\), while upper-right triangles use the transmission-eigenvalue spectra \(\{\tau_n(E)\}\). Large diagonal entries indicate disorder mechanisms whose spectra are assigned to their own class, whereas off-diagonal entries reveal disorder classes with similar conductance-balanced transport responses.
}
\label{fig:confusion_main}
\end{figure}

The distinguishability maps in Fig.~\ref{fig:confusion_main} show a structured pattern rather than a random distribution of errors. Bond disorder, dopants, and vacancy/edge disorder have strongly concentrated diagonal weight, indicating transport signatures that remain well separated after conductance balancing. The principal ambiguities occur among scalar-potential-like classes and the mixed class. Smooth onsite disorder, pixelated gates, Gaussian gates, and mixed perturbations can generate similar conductance-balanced spectral envelopes because they perturb related sectors of the Hamiltonian. By contrast, bond disorder modifies the hopping network, dopants introduce localized resonant scatterers, and vacancies or edge roughness disrupt lattice connectivity and boundary-sensitive modes. The off-diagonal entries therefore have a direct physical interpretation: they identify mechanisms that become close in the two-terminal transport representation, while the diagonal entries identify mechanisms that retain distinct spectral structure.

\subsection{Removing conductance-scale information}
\label{subsec:shape_controls}

\begin{figure}[t]
\includegraphics[width=\linewidth]{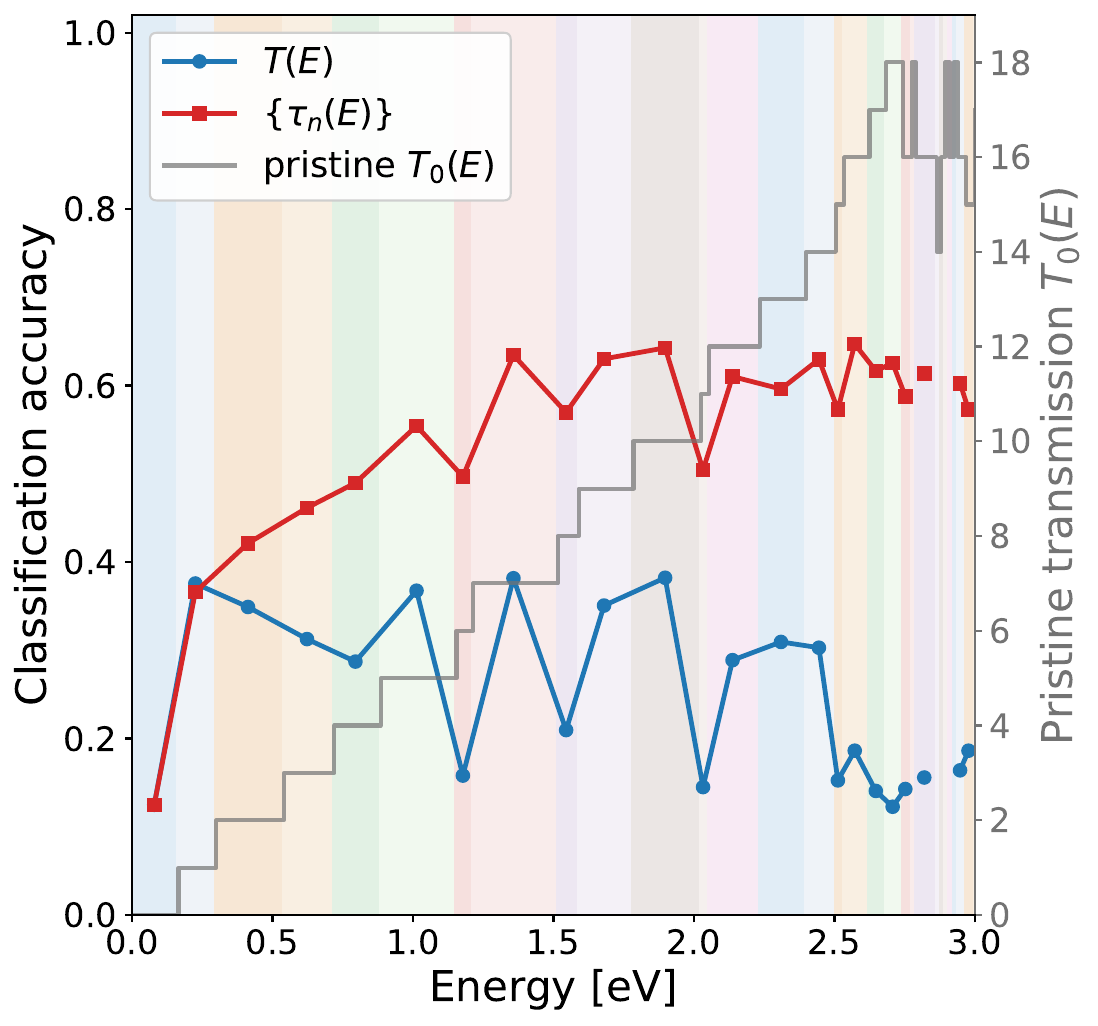}
\caption{
Spectral-shape controls for the conductance-balanced disorder ensemble. The figure compares disorder-class classification after transformations that suppress absolute conductance scale: raw spectra, area-normalized spectra, per-spectrum standardized spectra, and finite-difference spectra for the scalar transmission. Robust performance under area normalization and standardization indicates that the learned fingerprints are carried by spectral shape and channel redistribution rather than by average transmission.
}
\label{fig:shape_controls}
\end{figure}

The conductance-balanced construction already suppresses the strongest scale cue. A stricter test is to transform each spectrum before training. The scale-normalization controls are summarized in Fig.~\ref{fig:shape_controls}. For $T(E)$, the raw, area-normalized, and per-spectrum standardized inputs give classification accuracies of 0.883, 0.884, and 0.885, respectively. The scalar spectrum remains predictive after its absolute scale is removed. The derivative-only representation is weaker, with an accuracy of 0.666, indicating that local slopes alone capture only part of the fingerprint. The classifier therefore uses broader spectral structure: plateau distortions, correlations among energy windows, particle-hole asymmetry, and resonance patterns.

For the eigenchannel input, the corresponding raw, area-normalized, and standardized accuracies are 0.927, 0.915, and 0.921. The performance remains high under normalization. The recoverable information is therefore distinct from a residual conductance offset left by imperfect balancing. The persistence of class information in normalized eigenchannel spectra is one of the main pieces of evidence that disorder mechanisms imprint themselves through mode redistribution and energy-dependent scattering structure.

\subsection{Energy-window sensitivity}
\label{subsec:energy_windows}

By restricting the analysis to individual pristine-conductance plateaus, the energy-window tests reveal which mode sectors carry the strongest disorder-dependent structure. The classifier is trained on restricted spectral intervals and compared with the full-window result. The central low-energy region is the least diagnostic. Windows near the middle of the spectrum give classification accuracies near 0.63 for both scalar and channel-resolved inputs. This trend is consistent with the limited number of active propagating modes near the central sector of the armchair ribbon spectrum. When only a small set of modes contributes, different disorder mechanisms have fewer independent channel degrees of freedom through which to express distinct scattering signatures.

The outer multichannel regions are more informative. In the windows $[-3,-1.5]$ and $[1.5,3]$ eV, (see Fig.~\ref{fig:shape_controls} for the latter interval) the scalar spectrum reaches accuracies around 0.75, while the channel-resolved representation reaches approximately 0.81--0.82. The full spectrum remains stronger than any isolated window, with accuracies around 0.875 for $T(E)$ and 0.919 for the channel-resolved representation in the learning-curve analysis. The disorder fingerprint is therefore distributed across energy rather than localized at one diagnostic point. The high-energy sectors are important because multiple transverse modes contribute, allowing disorder to act through intermode scattering, selective channel suppression, and redistribution of transmission among open eigenchannels.

This result also clarifies the role of graphene in the study. The physical contrast arises less from a single graphene-specific band feature than from the coexistence of subband thresholds, transverse quantization, and lattice-scale scattering channels in a coherent waveguide. A different multichannel tight-binding system should display analogous behavior whenever its disorder mechanisms couple differently to propagating modes, boundary-sensitive states, and localized resonant structures.

The same analysis suggests a practical strategy for future calculations. If computation or measurement restricts the available energy range, windows with several open channels should be prioritized. Low-energy windows remain valuable for detecting band-edge shifts, localized midgap states, or strong vacancy-induced features, yet they provide fewer independent mode-mixing degrees of freedom. A broad multichannel window offers a richer response because each disorder mechanism can couple differently to transverse mode profiles, subband velocities, and boundary amplitudes.

\subsection{Channel redistribution behind the spectral fingerprints}
\label{subsec:channel_descriptors}

To connect the classification results with scattering physics, we analyze the distribution of transmitted current among eigenchannels using $N_{\rm eff}(E)$ and $S_\tau(E)$ from Eqs.~\eqref{eq:Neff_definition} and \eqref{eq:entropy_tau_definition}. These quantities summarize different aspects of the transmission-eigenvalue spectrum. $N_{\rm eff}$ measures the participation of eigenchannels in carrying current; $S_\tau$ measures how uniformly the transmitted current is distributed over channel space. Both quantities are insensitive to the detailed phase of scattering states, yet they are sensitive to whether disorder blocks a few channels, attenuates many channels, or redistributes transmission weight broadly.

\begin{figure}[t]
\centering
\includegraphics[width=\linewidth]{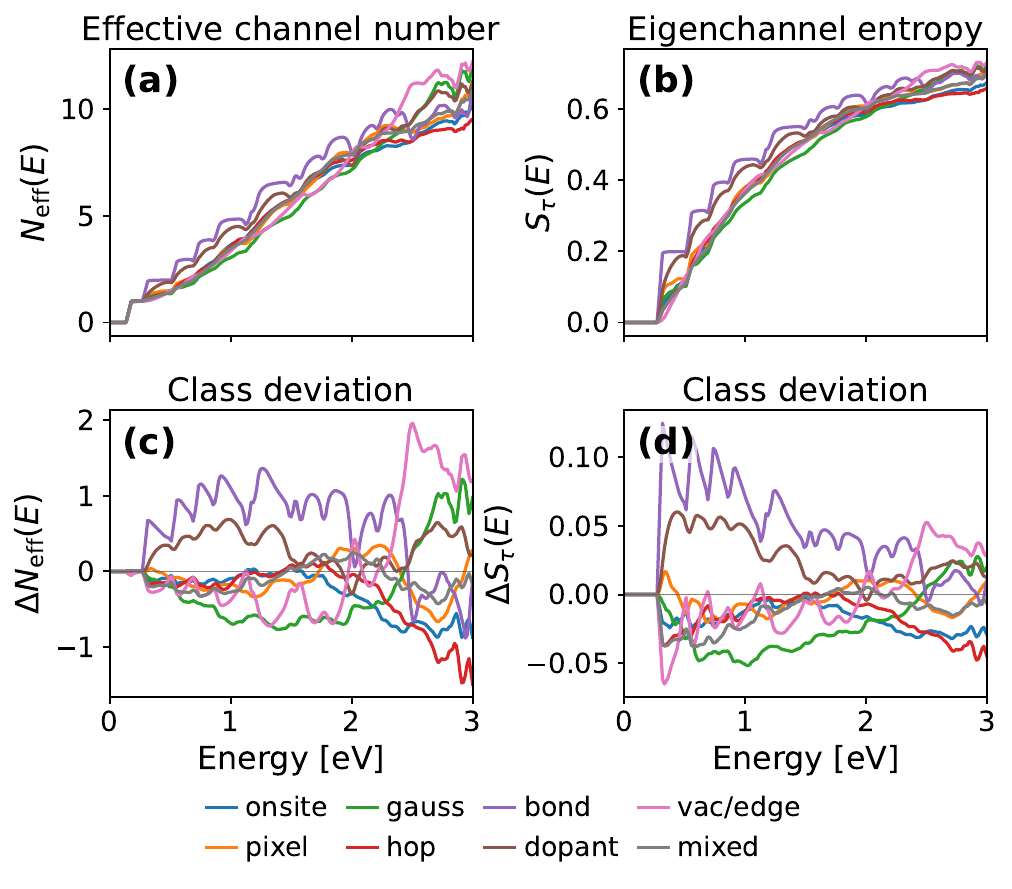}
\caption{Class-averaged eigenchannel descriptors in the conductance-balanced ensemble. (a) Effective number of transmitting eigenchannels $N_{\rm eff}(E)$. (b) Normalized eigenchannel entropy $S_\tau(E)$. (c,d) Deviations from the ensemble-averaged reference, which remove the common mode-opening envelope and emphasize class-dependent channel redistribution.}
\label{fig:channel_descriptors}
\end{figure}

Figure~\ref{fig:channel_descriptors} shows these descriptors for the conductance-balanced ensemble. Panels (a) and (b) display the class-averaged quantities $N_{\rm eff}(E)$ and $S_\tau(E)$. Their dominant increase with energy reflects the opening of additional propagating modes in the clean ribbon. To remove this common band-structure envelope and isolate class-dependent redistribution, panels (c) and (d) show
\begin{equation}
\begin{aligned}
\Delta N_{\rm eff}^{(c)}(E)&=
\langle N_{\rm eff}(E)\rangle_c
-
\langle N_{\rm eff}(E)\rangle_{\rm all},\\
\Delta S_\tau^{(c)}(E)&=
\langle S_\tau(E)\rangle_c
-
\langle S_\tau(E)\rangle_{\rm all},
\end{aligned}
\label{eq:descriptor_deviations}
\end{equation}
where $c$ labels the disorder class and $\langle\cdots\rangle_{\rm all}$ denotes the average over the full conductance-balanced ensemble.

The distinction between $N_{\rm eff}$ and $S_\tau$ is useful in interpreting these curves. $N_{\rm eff}$ is dominated by the participation of the largest eigenvalues and is therefore sensitive to whether the leading channels remain open. $S_\tau$ responds more directly to the relative uniformity of the eigenvalue distribution. A sample can have a moderate $N_{\rm eff}$ and a lower entropy if one channel remains dominant while several weak channels contribute small weights. Conversely, a more uniform set of partially transmitting eigenchannels raises the entropy. Reading the two descriptors together helps distinguish selective blocking from broad channel mixing.

The mean-subtracted descriptors reveal distinct channel-redistribution signatures. Bond disorder produces positive deviations in both $N_{\rm eff}$ and $S_\tau$ over broad energy intervals, consistent with random hopping perturbations that modify lattice connectivity and enhance transverse-mode mixing. Dopant disorder also increases channel participation in selected energy windows, with a different energy dependence that reflects localized resonant onsite scattering. Vacancy and edge disorder show stronger energy dependence, indicating that they can suppress selected channels or redistribute current among several partially transmitting modes depending on the available propagating states. Smoother scalar-potential disorder remains closer to the ensemble average over much of the spectrum, consistent with a coherent perturbation that is less efficient at randomizing the channel distribution.

These descriptors complement the full spectral classifiers by providing a compact physical interpretation of why disorder classes remain distinguishable after conductance balancing. Even when average transmission is controlled, different microscopic perturbations organize the transmitted current differently in eigenchannel space. $N_{\rm eff}$ and $S_\tau$ expose this mode-mixing and channel-participation structure directly.

\subsection{Learning curves and finite-sample stability}
\label{subsec:learning_curves}

Learning curves are used as a stability check. The classification accuracy increases rapidly when the training set grows from 5\% to 20\% and then approaches a plateau. For $T(E)$ the full-training accuracy is approximately 0.88 in the learning-curve run; for the channel-resolved representation it is approximately 0.92. The regression metrics follow the same qualitative trend. This behavior indicates that the reported fingerprints are robust across different train/test splits. Since the classification curves are already close to saturation, increasing the data set should primarily refine regression, confidence intervals, and class-boundary structure rather than change the qualitative conclusion.

The finite-sample analysis also clarifies the status of the mixed class. Its overlap with scalar-potential-like classes is more than failure of the classifier. Mixed disorder occupies intermediate regions of transport space because its Hamiltonian contains several scattering mechanisms at once. Additional samples would sharpen the boundaries of that region, but the overlap itself reflects the physics of superposed perturbations.

\section{Discussion}
\label{sec:discussion}

The results support a metrological interpretation of inverse quantum transport. A deterministic map from a scalar spectrum to an atom-resolved disorder field is an ill-suited target for a two-terminal coherent measurement, because the scattering map compresses a high-dimensional Hamiltonian into a finite set of boundary observables. The conductance-balanced benchmark instead asks which coarse Hamiltonian information remains recoverable once average transmission is controlled. For the disorder families studied here, the answer is affirmative: transport spectra retain mechanism-dependent information in their energy dependence and in the distribution of transmission among eigenchannels.

The confusion structure is part of this conclusion. Smooth onsite disorder, pixelated gates, Gaussian gates, and mixed disorder overlap because several of them act primarily through scalar onsite potentials. Bond disorder, dopants, and vacancy/edge disorder leave more distinct fingerprints because they perturb connectivity, introduce localized resonances, or disrupt boundary-sensitive modes. The energy-window analysis adds a complementary interpretation: multichannel regions are more diagnostic than the central low-energy sector, consistent with the idea that disorder mechanisms separate most clearly when several propagating modes are available to be mixed, suppressed, or redistributed.

The channel-resolved results should be interpreted as a hierarchy of observable information. In numerical transport, the full matrix $t(E)$ and its eigenvalues are directly available. In an experiment, a conventional two-terminal conductance measurement gives only $\sum_n\tau_n$. Shot noise gives access to $\sum_n\tau_n[1-\tau_n]$, and higher current cumulants or multi-terminal/tomographic protocols would be needed to reconstruct more of the eigenvalue distribution. The present work therefore establishes the information content of the transport response under idealized coherent conditions. It also identifies which additional measurements would be most valuable experimentally: observables sensitive to the variance and higher moments of the transmission eigenvalues should distinguish disorder mechanisms more effectively than conductance alone.

The framework goes beyond graphene-specific band-structure features, but to the availability of a coherent scattering description for a finite region coupled to external channels~\cite{Datta1995,BeenakkerRMP1997,Groth2014Kwant}. The armchair ribbon is used here as a controlled tight-binding waveguide with atomically defined edges, transverse subbands, and several independently tunable disorder channels~\cite{BreyFertig2006,Tworzydlo2006,Groth2014Kwant}. The same protocol can be transferred to semiconductor quantum wires and molecular tight-binding junctions, where conductance is naturally formulated in terms of channel transmission or Green-function scattering approaches~\cite{vanWees1988,Wharam1988,NitzanRatner2003,Datta1995}. It can also be adapted to superconducting hybrid structures, spinful chains, and photonic or acoustic scattering networks, provided the measured response can be represented by an appropriate normal, Bogoliubov--de Gennes, spin-resolved, or classical-wave scattering matrix~\cite{BlonderTinkhamKlapwijk1982,BeenakkerVanHouten1991,DattaDas1990,ZhangMonticoneMiller2023,ZhangDelaporteMa2025}. The general principle is that inverse transport should target the part of the Hamiltonian disorder that is recoverable from the chosen observable, since distinct Hamiltonians or graph structures can produce identical or incomplete scattering data when the scattering map contains symmetries, projections, or other degeneracies~\cite{BomanKurasov2005,BlastenExnerIsozakiLassasLu2024}.

Two methodological limitations should be kept explicit. First, PCA and random forests operate on sorted eigenvalue ranks, providing a representation of spectral ordering rather than continuous tracking of individual transverse modes through avoided crossings or subband openings. This representation is appropriate for transmission eigenvalue statistics, yet it discards phase information and the detailed composition of the eigenvectors in lead-mode space. Second, the class labels are generated by simulation priors. A real sample may contain disorder outside this catalog, and its spectrum would then be projected onto the closest learned transport class. These limitations clarify the scope within which the passive-metrology result should be applied.

The controlled coherent-scattering baseline established here opens a route toward a broader program of disorder metrology.  The extensions listed below therefore serve to transfer and stress-test the fingerprinting protocol across additional physical regimes, while the present work defines the reference limit in which the disorder signatures are cleanly isolated.
Varying ribbon width, length, edge termination, and lead geometry would separate signatures that are generic to multichannel waveguides from those tied to the armchair realization, whose subbands, boundary conditions, edge disorder, and contact geometry are known to shape graphene-ribbon transport~\cite{BreyFertig2006,Tworzydlo2006,LibischRotterBurgdorfer2012,Stegmann2017ContactGeometry}. Finite temperature, dephasing, electron--electron interactions, and electron--phonon scattering can be introduced as controlled broadening and relaxation mechanisms, allowing the crossover from ideal coherent spectra to experimentally realistic device responses to be quantified~\cite{Datta1995,FoersterSamuelssonPilgramBuettiker2007,MeirWingreen1992,FrederiksenPaulssonBrandbygeJauho2007}. Relaxing the ideal-contact assumption by adding interface disorder and imperfect mode matching would place contact-induced spectral structure on the same footing as intrinsic disorder, which is important because electrode coupling and contact geometry can strongly affect nanoscale conductance spectra~\cite{Stegmann2017ContactGeometry,BrandbygeMozosOrdejonTaylorStokbro2002}. Finally, sweeping the disorder-strength priors within each class would distinguish confusion patterns that are stable properties of the scattering mechanisms from those that reflect a particular sampling distribution~\cite{HastieTibshiraniFriedman2009,QuioneroCandela2008DatasetShift}.
Fifth, a direct bridge to experimentally accessible observables can be developed by extending the present two-terminal conductance analysis to additional response functions. Shot noise is a natural first extension because it probes transmission partitioning and is directly sensitive to the distribution of transmission eigenvalues~\cite{BlanterButtiker2000,Kumar1996ShotNoise,Reznikov1995ShotNoise}. Thermopower and thermal conductance provide complementary access to the energy dependence of the transmission spectrum~\cite{SivanImry1986Thermopower,Butcher1990Thermoelectric,Molenkamp1990ThermopowerQPC}. Finite-bias spectroscopy and gate-dependent conductance maps are standard probes of mesoscopic and molecular junctions, and would test whether the disorder signatures identified here persist beyond linear response~\cite{MeirWingreen1992,NitzanRatner2003}. Magnetotransport and gate-modulated response functions add phase- and interference-sensitive dimensions to the scattering data, providing further handles on disorder-dependent transport structure~\cite{LeeStone1985UCF,Marcus1992ConductanceFluctuations,Alhassid2000QuantumDots}.

\section{Conclusions}
\label{sec:conclusions}

We have formulated inverse quantum transport as a conductance-controlled disorder-metrology problem and tested it in ensembles of disordered armchair graphene nanoribbons. The benchmark is balanced by disorder class and by windows of energy-averaged transmission, preventing the task from reducing to a simple ordering by conductance loss. Within this controlled ensemble, transport spectra retain mechanism-specific fingerprints. These fingerprints survive area normalization and per-spectrum standardization, and energy-window analysis shows that they are strongest in multichannel spectral regions.

The physical origin of the fingerprints is channel redistribution. Scalar transmission records only the total transmitted probability, while the transmission-eigenvalue spectrum resolves how current is partitioned among eigenchannels. Disorder mechanisms that look similar in average conductance can differ in their mode mixing, selective channel suppression, resonance structure, and boundary sensitivity. PCA transport coordinates and eigenchannel participation descriptors provide complementary views of this structure. PCA identifies the dominant spectral deformation directions, and $N_{\rm eff}$ together with $S_\tau$ connects those directions to current redistribution among channels.

The introduction posed the central question of what can be inferred about a high-dimensional disordered Hamiltonian from finite coherent transport observables. The conclusion is deliberately metrological.  It is unlike that two-terminal spectra could provide a unique image of the microscopic disorder realization. They nevertheless encode reproducible information about the mechanism and statistical character of the disorder, provided that average conductance is controlled and the channel-resolved structure of the scattering response is considered.  This result provides a physically interpretable basis for Hamiltonian disorder metrology in coherent quantum waveguides. It also establishes a reference problem from which a continuing sequence of studies can add geometric variation, contact nonideality, finite temperature, dephasing, interactions, phonons, finite-bias driving, magnetotransport, and active gate probes in a controlled order.

\section{Acknowledgments}
Los Alamos National Laboratory is managed by Triad National Security, LLC, for the National Nuclear Security Administration of the U.S. Department of Energy under Contract No. 89233218CNA000001.

\setcitestyle{notesep={;}}
\bibliography{biblio}
\end{document}